\documentclass[10pt,article]{IEEEtran}
\usepackage{amssymb}
\usepackage{amsmath}

\title{Coherence Analysis of Iterative Thresholding Algorithms}

\author{
Arian Maleki\\
Department of Electrical Engineering and Statistics, \\
Stanford University\\
arianm@stanford.edu
}

\newtheorem{thm}{Theorem}[section]
\newtheorem{lem}[thm]{Lemma}
\begin{document}

\maketitle
\begin{abstract}
There is a recent surge of interest in developing algorithms for finding sparse solutions of underdetermined systems of linear equations $y = \Phi x$. In many applications, extremely large problem sizes are envisioned, with at least tens of thousands of equations and hundreds of thousands of unknowns. For such problem sizes, low computational complexity is paramount. The best studied $\ell_1$ minimization algorithm is not fast enough to fulfill this need. Iterative thresholding algorithms have been proposed to address this problem. In this paper we want to analyze two of these algorithms theoretically, and give sufficient conditions under which they recover the sparsest solution.
\end{abstract}
\section{introduction}
Finding the sparsest solution of an underdetermined system of linear equations $y = \Phi x$, is a problem of interest in signal
processing, data transmission, biology and statistics-just to name a few. Unfortunately, this problem is NP-hard and in general
can not be solved by a polynomial time algorithm. Chen et al. \cite{ChDoSa98} proposed the following optimization for recovering the sparsest solution;
\begin{align} \label{ell1}
(\mathcal{Q}_1) \ \ \ \min \|x\|_1 \ \ \ \ \  \textmd{s.t.} \ \Phi x=y, \nonumber
\end{align}
where $\ell_p$-norm is defined as $\|x\|_p= \sqrt[p]{\sum_i{|x_i|^p}}$. \\
Greedy methods have also been proposed as another alternative for solving such a problem. One of the best known algorithms of this class is orthogonal matching pursuit (OMP) \cite{PaReKr93}. Intuitively speaking at each iteration, OMP finds a column of $\Phi$ which has the maximum correlation with the error of the approximation up to this step, and adds it to the active set and projects $y$ onto the range of the active set to get a new estimate. The third class of algorithms that has drawn a lot of attention recently is the class of iterative thresholding algorithms. This class has the least computational complexity and is the most suitable class for very large scale problems \cite{MaDo09}. There are many theoretical results that prove the optimality of the first two classes of algorithms under certain conditions, but there are much less rigorous results for thresholding algorithms. Before mentioning some of the results, we first set up the notation we are going to use in the paper.
Suppose that $x_o \in \mathbb{R}^N$ is a $k$ sparse vector (i.e.~it has at most $k$ non-zero elements). We observe the measurement vector $y= \Phi x_o$ which is in $\mathbb{R}^n$ ($n< N$) and the goal is to reconstruct the original vector $x_o$. Without loss of generality, we assume that the columns of $\Phi$ have unit $\ell_2$ norm. Another notation that is used in the paper is the notion of restricted submatrices. For a subset of columns of $\Phi$ called $J$, $\Phi_J$ includes all the columns of $\Phi$ whose indices are in $J$, and $x_J$ all the elements of $x$ whose indices are in $J$. The coherence of $\Phi$ is defined as,
\begin{align}
\mu= \max_{\{i,j: 1 \leq i,j \leq N, i \neq j\}} |\langle \phi_i, \phi_j \rangle |.
\end{align}
where $\phi_i$ is the $i^{\rm th}$ column of the matrix $\Phi$.
In the following, a summary of the results proved for $\ell_1$ minimization and OMP algorithms in \cite{DoEl03}  and \cite{Tr04} respectively, are presented.

\begin{thm}
If $k \leq \frac{1}{2}(1+ \mu^{-1})$, then both the $\ell_1$ minimization and the OMP recover the sparsest solution.
\end{thm}

When the matrix $\Phi$ is drawn from a random ensemble \cite{Donoho1,CaRoTa06}, we can bound the coherence \cite{Gilbert_OMP}, and find conditions for the exact sparse signal recovery. In this random setting, however, the results can be improved \cite{DoTa09}.
Although the theoretical results are basically focused on $\ell_1$ relaxation and greedy methods, many large scale applications have already moved toward the thresholding algorithms \cite{FiNo03}, \cite{StElDo05}. In a recent paper, we considered a few thresholding policies and showed that the results of these algorithms are very impressive in practical situations such as compressed sensing \cite{MaDo09}. In this paper we focus on the  theoretical aspects of these algorithms. \\
The organization of the paper is as follows. In Section \ref{sec:ITA}, we discuss the thresholding algorithms and the thresholding policy considered in the paper; The main results of the paper will also be reviewed. Section \ref{sec:proof1} presents the convergence proof of the thresholding algorithms. In Section \ref{discussion}, we will briefly review the existing literature on iterative thresholding algorithms and compare those results to ours. Finally Section \ref{conclusion} concludes the paper.

\section{Iterative Thresholding Algorithms}\label{sec:ITA}

\subsection{Abstracted thresholding Algorithm}
Consider two threshold functions $\eta_t(x)$ to be applied elementwise to vectors: hard thresholding $\eta_{\mu}^H(x)= x\mathbf{1}_{\{ |x| > \mu\}}$ and soft thresholding $\eta_{\mu}^S(x)= {\rm sgn}(x) (|x|- \mu)_+$, where $\mathbf{1}$ is the indicator function and $(a)_+$ is equal to $a$ if $a>0$, and zero otherwise. Iterative hard thresholding (IHT) and iterative soft thresholding (IST) algorithms are defined with the following iteration,
\begin{align}\label{ihst_eq}
x^{t+1}= \eta_{\lambda_t}^*(x^t+ \Phi^T (y- \Phi x^t)),
\end{align}
where $\lambda_t$ is the threshold value at time $t$, $* \in \{H,S\}$ represents hard or soft thresholding, $\Phi^T$ is the transpose of the matrix $\Phi$ and $x^t$ is our estimate at time $t$. Note that the threshold value may depend on the iteration. The basic intuition is that since the solution  satisfies the equation $y= \Phi x$, algorithm makes progress by moving in the direction of the gradient of $\|y-\Phi x\|^2$ and then by thresholding the result, it tries to get a sparse vector closer to the hyperplane $y=\Phi x$. Another intuition for this algorithm comes from \cite{DaDeDe04} and is as follows. Suppose that we want to solve the following optimization problem,
\begin{equation} \label{pq}
(\mathcal{P}_q) \ \ \ \ \  \min_x  \  \|y-\Phi x\|_2^2+ 2\lambda \|x\|_q. \nonumber
\end{equation}
It has been proved that the following IST algorithm converges to the solution of $\mathcal{P}_1$ in case $\|\Phi^T\Phi-I\|_{2,2}<1$,
\begin{align}\label{fixedthresh}
x^{t+1}= \eta_{\lambda}^S(x^t+ \Phi^T (y- \Phi x^t)),
\end{align}
where $\|A\|_{2,2}$ is the spectral norm of the matrix $A$. It may be noted that $\lambda$ is fixed here and does not depend on the iteration.
It is also well-known that as $\lambda \rightarrow 0$ the solution of $\mathcal{P}_1$ converges to the solution of $\mathcal{Q}_1$. But it is easy to see if $\Phi$ is a fat matrix, setting $\lambda$ to a very small value in  \eqref{fixedthresh} will not work. A proper thresholding policy is to set the threshold to a large value and gradually decrease it as the algorithm proceeds. The following theorem justifies this intuition.

Consider the iterative soft or hard thresholding algorithms introduced in equation (\ref{ihst_eq}). Suppose  $\lambda_t \rightarrow 0$ as $t \rightarrow \infty$,  and $\lambda_t$ is a decreasing sequence (this condition may not hold, but for the simplicity of the proof we assume it is true). Let $J_t$ denote the union of the support of $x^t$ and $x_o$ and define $L_t := J_{t+1} \cup J_{t}$.  Assume that $L_t$ satisfies, $\sup_t \|I-\Phi^T_{L_t}\Phi_{L_t} \|_{2,2}=\gamma <1$. Under these conditions:
\begin{thm}
The iterative thresholding algorithm will converge to the sparsest solution.
\end{thm}
\begin{proof}
\begin{align}
& \|x^{t+1}-x_o\|_2= \|x^{t+1}_{L_{t+1}}-x_{o_{L_{t+1}}}\|_2  \nonumber \\
&\quad\quad\leq \|\eta_{\lambda_{t+1}}^{*}(x^t_{L_t}+ \Phi_{L_t}^T(\Phi_{L_t} x_{o_{L_t}}-\Phi_{L_t} x^t_{L_t}))-x_{o_{L_t}}\|_2,\nonumber \\
&\quad\quad \leq \|(x^t_{L_t}+ \Phi_{L_t}^T(\Phi_{L_t} x_{o_{L_t}}+ -\Phi_{L_t} x^t_{L_t}))+ \epsilon_{t+1} -x_{o_{L_t}}\|_2,\nonumber \\
&\quad\quad \overset{1}{\leq} \|(I-\Phi_{L_t}^T \Phi_{L_t})(x^t_{L_t}-x_{o_{L_t}})\|_2 +\sqrt{n}\lambda_{t+1}, \nonumber\\
&\quad\quad \leq \|(I-\Phi_{L_t}^T \Phi_{L_t})\|_{2,2} \|x^t_{L_t}-x_{o_{L_t}}\|_2+\sqrt{n}\lambda_{t+1}, \nonumber
\end{align}
where $\epsilon_{t+1}$ is an extra error introduced by the thresholding process and therefore each element of this vector is less than $\lambda_{t+1}$. Also all the elements that are not in $L_t$ are zero. Inequality (1) is just the triangle inequality for $\ell_2$ norm.
For any $\epsilon>0$, choose $T_0$ such that $\sqrt{n}\lambda_{T_0+1} < \frac{\epsilon(1-\gamma)}{2}$, and let $\|x^{T_0+1}-x_o\|_2 = e $. Then, find $T_1$ such that $\gamma^{T_1}e < \epsilon/2$. Now it is easy to prove that at $t=T_0+T_1$, the error is less than $\epsilon$ and therefore the total error goes to zero.
\end{proof}

This theorem is not useful for practical purposes since we should have information on the size of $L_t$. In the next section we mention a practical thresholding policy that may be used in practice.

\subsection{Thresholding Policy}
 Suppose that an oracle tells us the true underlying $k$. Then since the final solution is $k$ sparse, the threshold can be set to the magnitude of the $(k+1)^{\rm th}$ largest coefficient. This type of thresholding policy has also been used in \cite{NeTr08},\cite{DaMi09}, \cite{BlDa09}.
  The only problem is how to get the oracle information. In a recent paper, we showed how one can de-oraclize such algorithms for compressed sensing problems \cite{MaDo09}. For other types of problems, $k$ may be estimated using cross validation. If neither of these two methods is applicable, the bounds derived in this paper for the sparsity may be used for setting $k$. From now on, whenever we refer to IHT or IST, the thresholding policy is the $k$ largest element thresholding policy unless otherwise stated.

\subsection{Main Results}
We will prove two main theorems for the two thresholding algorithms that have been mentioned in the last section.

\begin{thm} \label{hard_thm}
Suppose that $k<\frac{1}{3.1}\mu^{-1}$ and $ \frac{|x_o(i)|}{|x_o(i+1)|}< 3^{\ell_i-4} , \forall i,\ {1 \leq i <k}$.  Then IHT finds the correct active set in at most $\sum_{i=1}^k \ell_i+k $ steps. After this step all of these elements will remain in the active set and the error will go to zero exponentially fast.
\end{thm}

\begin{thm} \label{soft_thm}
Suppose that $k<\frac{1}{4.1}\mu^{-1}$ and $\forall i,\ {1 \leq i <k}$, we have $\frac{|x_o(i)|}{|x_o(i+1)|}< 2^{\ell_i-5} $.  Then IST recovers the correct active set in at most $\sum_{i=1}^k  \ell_i+k $ steps. After that all these coefficients will remain in the active set and the error will go to zero exponentially fast.
\end{thm}

 The sufficient conditions provided here are slightly weaker than the conditions mentioned for $\ell_1$ or OMP. Simulation results also confirm that IHT and IST are weaker than $\ell_1$ in practice \cite{MaDo09}. Another interesting fact is that the number of iterations needed, depends on the ratio of the coefficients but this dependency is roughly logarithmic and therefore it will work well in practice. Also, the algorithms find the correct active set in a finite number of iterations and once the algorithms find the correct active set, they converge to the exact solution exponentially fast.

\section{Proof of Convergence for the IHT Algorithm}\label{sec:proof1}
The goal of this section is to give an outline of the proof of Theorem \ref{hard_thm}. We define the following two variables,
\begin{align}
z^i&= x^i + \Phi^T (\Phi x_o- \Phi x^i), \\
w^i&= x_o-x^i,
\end{align}
where $x_o$ is the optimal value and $x^i$ is our estimate at the $i^{\rm th}$ step. The $j^{\rm th}$ element of these two vectors will be denoted by $z^i(j)$ and $w^i(j)$. The active set of $x^i$ is called  $I^{i}$. Finally, $x_o(i)$ denotes the $i^{\rm th}$ element of $x_o$. Without loss of generality we assume that $x_o(i)$'s are sorted in descending order of their absolute values and therefore the only non-zero elements of $x_o$ are the first $k$ elements. The next lemma will be useful later when we try to bound the error at each iteration.
\begin{lem}\label{seq_lem}
Consider the following sequence for $s \geq 0$,
\begin{align}
f_s= \alpha^1 + \ldots \alpha^s +\beta \alpha^{s+1}, \nonumber
\end{align}
where $0<\alpha <1$. The following statements are true;
\begin{enumerate}
  \item If $\beta(1-\alpha) <1$, then for every $s$, $f_s < \frac{\alpha}{1-\alpha}$.
  \item If $\beta(1-\alpha) >1$, then for every $s$, $f_s < \beta \alpha$.
  \item If $\beta(1-\alpha)=1$, then $f_s$ is a constant sequence and is always equal to $\frac{\alpha}{1-\alpha}$.
\end{enumerate}
\end{lem}
It is easy to see that the sequence is either increasing or decreasing or constant depending on the values of $\alpha$ and $\beta$. The proof is simple and is omitted for the sake of brevity.

\begin{lem} \label{hard_lem_2}
Suppose that $x_o(1),x_o(2), \ldots, x_o(r-1)$, $r-1 < k$,  are in the active set at the $m^{\rm th}$ step. Also assume that,
\begin{align}
|z^m(j)-x_o(j)| \leq 1.5 k\mu |x_o(r-1)| \ \ \  \ \forall j. \nonumber
\end{align}
If $k\mu < \frac{1}{3.1}$, then at stage $m+ s$ and for every $j$ we will have the following upper bound for $|z^{m+s}(j)-x_o(j)|$,
\begin{align} \label{mainlemma_hard}
|x_o(r)|\left( k\mu + \ldots +(k\mu)^{s}\right)+1.5(k\mu)^{s+1}|x_o(r-1)|.
\end{align}
Moreover, $x_o(1),x_o(2), \ldots, x_o(r-1)$ will remain in the active.
\end{lem}
\begin{proof}
We prove this by induction; Assuming that the bound holds at stage $m+s$ and $x_o(1),x_o(2), \ldots, x_o(r-1)$ are in the active set, we show that the upper bound holds at stage $m+s+1$ and the first $r-1$ elements will remain in the active set.
 \begin{align}
 &|z^{m+s+1}(i)-x_o(i)| \nonumber \\
& \leq  \sum_{j \in I^{m+s} \backslash \{i\}} \! \! \! \! \! |\langle \phi_i, \phi_j \rangle w^{m+s}(j)| + \hspace{-0.8cm}\sum_{\tiny j \in \{1,2, \ldots k\} \backslash I^{m+s}\cup \{i\}}\! \! \! \! \! \! \! \! \! \! \! \! |\langle \phi_i, \phi_j \rangle  w^{m+s}(j)|, \nonumber\\
  &\overset{1}{=}\sum_{j \in I^{m+s} \backslash \{i\}} \! \! \! \! \! |\langle \phi_i, \phi_j \rangle w^{m+s}(j)| + \hspace{-0.8cm}\sum_{\tiny j \in \{r, \ldots k\} \backslash I^{m+s}\cup \{i\}}\! \! \! \! \! \! \! \! \! \! \! \! |\langle \phi_i, \phi_j \rangle  w^{m+s}(j)|, \nonumber\\
  &\overset{2}{\leq} \sum_{j \in I^{m+s} \backslash \{i\}}|\langle \phi_i, \phi_j \rangle (z^{m+s}(j)-x_o(j))| + k\mu x_o(r), \nonumber\\
  &\leq k\mu|x_o(r)| (k\mu + \ldots + (k\mu)^{s}) + 1.5(k\mu)^{s+2}|x_o(r-1)|\nonumber \\
  &\ \ \ \ +k\mu |x_o(r)|,   \nonumber \\
  &\leq |x_o(r)|(k\mu + \ldots + (k\mu)^{s+1}) + 1.5(k\mu)^{s+2}|x_o(r-1)|. \nonumber
 \end{align}
In these calculations equality (1) is due to the assumptions of the induction, i.e. the first $r-1$ elements are in the active set at stage $m+s$. To get inequality (2) we have used two different facts. The first one is that when $j \in I^{m+s}$, $w^{m+s}(j)=x_o(j)-z^{m+s}(j)$ and the second one is that when $j \in \{r, \ldots k\} \backslash I^{m+s}$ then $w^{m+s}(j)=x_o(j)$ and therefore $|x_o(j)| \leq |x_o(r)|$. The last step is to prove that all the first $r-1$ elements remain in the active set. For $i \in \{1,2 \ldots, r-1\}$,

\begin{align}
	 |z^{m+s+1}(i)| & \geq |x_o(i)|- |z^{m+s+1}(i)-x_o(i)|, \nonumber \\
 	\overset{1}{\geq}  |x_o(i)| & -(k\mu |x_o(r-1)| + \ldots +(k\mu)^{s+1}|x_o(r-1)|) \nonumber \\
      -1.5  ( k\mu&)^{s+2}|x_o(r-1)| \overset{2}{\geq} |x_o(i)|- \frac{|x_o(r-1)|}{2.05} \nonumber\\
      \geq |x_o(r&-1)|-\frac{|x_o(r-1)|}{2.05}. \nonumber
\end{align}
In inequality (1) we have used the bound in (\ref{mainlemma_hard}) by replacing $x_o(r)$ with $x_o(r-1)$. Inequality (2) is the result of Lemma \ref{seq_lem}.
For $i \notin \{ 1,2 \ldots k \}$, we have
\begin{align}
	 |z^{m+s+1}(i)| \leq \frac{|x_o(r-1)|}{2.05}, \nonumber
\end{align}
and since $\min_{\{i: i \leq r-1\}}|z^{m+s+1}(i)| > \max_{\{i: i > k\}} |z^{m+s+1}(i)|$, the first $r-1$ elements will remain in the active set. The base of the induction is the same as the assumptions of this lemma and the proof is complete.
\end{proof}

\begin{lem} \label{hard_lem_3}
Suppose that $k<\frac{1}{3.1}\mu^{-1}$, and $x_o(1),x_o(2), \ldots, x_o(r)$, $r < k$,  are in the active set at the $m^{\rm th}$ step. Also assume that $\frac{|x_o(r)|}{|x_o(r+1)|} \leq 3^{\ell_r-4}$. If
\begin{align}
|z^m(j)-x_o(j)| \leq 1.5k\mu |x_o(r)| \ \ \  \ \forall \ j, \nonumber
\end{align}
after $\ell_r$ more steps $x_o(r+1)$ will get into the active set, and
\begin{align}
|z^{m+ \ell_r+1}(j)-x_o(j)| \leq 1.5k\mu |x_o(r+1)| \ \ \ \ \forall \ j. \nonumber
\end{align}
\end{lem}
\begin{proof} By setting $q= \ell_r$ in the upper bound we get,
\begin{align}
&|z^{m+\ell_r}(j)-x_o(j)| \leq \frac{1.5|x_o(r+1)|}{273}+ \frac{|x_o(r+1)|}{2.1}. \nonumber
\end{align}
Similar to the last lemma it is also not difficult to see that
\begin{align}
|z^{m+ \ell_r}(r+1)|& = |z^{m+ \ell_r}(r+1) -x_o(r+1) +x_o(r+1) | \nonumber \\
&\geq |x_o(r+1)|- |z^{m+ \ell_r}(r+1) -x_o(r+1) |  \nonumber \\
& \geq |x_o(r+1)|- \frac{|1.5x_o(r+1)|}{273}- \frac{|x_o(r+1)|}{2.1}. \nonumber
\end{align}
But,
\begin{equation}
 |z^{m+ \ell_r}(r+1)| > \max_{\{i: i>k\}}|z^{m+ \ell_r}(i)|, \nonumber
\end{equation}
and therefore $x_o(r+1)$ will be detected at this step. It may also be noted that at this stage the error is less than $|x_o(r+1)|/2$. For the next stage we will have at most $k $ active elements the error of each is less than $|x_o(r+1)|/2$ and at most $k-r$ non-zero elements of $x_o$ that have not passed the threshold and whose magnitudes are smaller than $|x_o(r+1)|$. Therefore, the error of the next step is less than $1.5k\mu |x_o(r+1)|$.
\end{proof}
Our goal is to prove the correctness of IHT by induction and we have to know the correctness of IHT at the first stage. The following lemma provides this missing step.
\begin{lem} \label{hard_lem_1}
Suppose that $k<\frac{1}{3.1}\mu^{-1}$, then at the first stage of the IHT, $x_o(1)$ will be in the active set\footnote{This result holds even if $k \mu <\frac{1}{2}$. For the sake of consistency with the other parts of the proof we state it in this way} and $|z^1(j)-x_o(j)| \leq k\mu |x_o(1)|$.
\end{lem}
\begin{proof}
\begin{align}
|z^1(1)|& \geq |x_o(1)|- k\mu |x_o(1)|. \nonumber
\end{align}
On the other,
\begin{equation}
\max_{\{i:k<i\}} |z^1(i)|= \max_{\{i:k<i\}} |\sum_{j=1}^k \langle \phi_i, \phi_j \rangle x_o(j)| \leq k \mu |x_o(1)|. \nonumber
\end{equation}
Therefore, since $k\mu<1-k\mu$, the index of the first element will be in the active set after the first step. The last claim of the Lemma is also clear.
\end{proof}
Finally the following lemma describes the performance of the algorithm after detecting all the non-zero elements.
\begin{lem} \label{hard_lem_4}
Suppose that $x_o(1),x_o(2), \ldots, x_o(k)$, are in the active set at the $m^{\rm th}$ step. Also assume that,
\begin{align}
|z^m(j)-x_o(j)| \leq 1.5 k\mu |x_o(k)| \ \ \  \ \forall j. \nonumber
\end{align}
If $k\mu < \frac{1}{3.1}$, then at stage $m+ s$ and for every $j$ we will have,

\begin{align}
|z^{m+s}(j)-x_o(j)| \leq 1.5(k\mu)^{s+1}|x_o(k)|. \nonumber
\end{align}
\end{lem}

Since the proof of this lemma is very similar to the proof of Lemma \ref{hard_lem_2}, it is omitted.
\begin{proof}[Outline of the proof of Theorem \ref{hard_thm}]
The proof is an induction that combines the above lemmas. Suppose that $x_o(1), x_o(2), \ldots, x_o(r)$ are already in the active set. According to Lemma \ref{hard_lem_2} all these terms will remain in the active set, and according to Lemma \ref{hard_lem_3} after ${\ell_r} $ steps $x_o(r+1)$ will also get into the active set. In one more step, the error on each element gets smaller than $1.5 k\mu |x_o(r+1)|$, and everything can be repeated. Lemma \ref{hard_lem_1} provides the first step of the induction. Finally when all the elements are in the active set lemma \ref{hard_lem_4} tells us that the error goes to zero exponentially fast.
\end{proof}

Since the proof of the convergence of IST is very similar to IHT we do not repeat it here. You may refer to \cite{Arian09} for more details.

\section{Proof of convergence for the IST algorithm}\label{sec:proof2}
As mentioned before the main ideas of the proof of the IST algorithm are very similar to those of the IHT. We will mention the proof in detail but will try to emphasize more on the differences. The following lemma helps us find some bounds on the error of the algorithm at each step.

\begin{lem}\label{lem_soft_1}
Suppose that $x_o(1), x_o(2), \ldots, x_o(r)$, $r \leq k$, are in the active set at the $m^{\rm th}$ step. Also assume that
\begin{align}
|x^{m}(j)-x_o(j)| \leq 4k\mu |x_o(r)|,   \ \ \ \ \forall j \in I^{m}, \nonumber
\end{align}
and $k\mu < \frac{1}{4.1}$. Then at stage $m+s$, $\forall \ i \in I^{m+s}$ we have the following upper bound for $|x^{m+s}(i)-x_o(i)|$,
\begin{align}
|x_o(r+1)|&\left(2k\mu + \ldots + (2k\mu)^{s}\right)+2(2k\mu)^{s+1}|x_o(r)|. \nonumber
\end{align}
Moreover, $x_o(1), x_o(2), \ldots, x_o(r)$ remain in the active set.
\end{lem}
\begin{proof}
As before, this can be proved by induction. We assume that at step $m+s$ the upper bound holds and $x_o(1), x_o(2), \ldots, x_o(r)$ are in the active set and we prove the same things for $m+s+1$. Similar to what we saw before,
\begin{align} \label{equation_soft}
&|z^{m+s+1}(i)-x_o(i)| \nonumber \\
& \leq  \sum_{j \in I^{m+s} \backslash \{i\}} \! \! \! \! \! |\langle \phi_i, \phi_j \rangle w^{m+s}(j)| + \hspace{-0.8cm}\sum_{\tiny j \in \{1,2, \ldots k\} \backslash I^{m+s}\cup \{i\}}\! \! \! \! \! \! \! \! \! \! \! \! |\langle \phi_i, \phi_j \rangle  w^{m+s}(j)|, \nonumber\\
  &\overset{1}{=}\sum_{j \in I^{m+s} \backslash \{i\}} \! \! \! \! \! |\langle \phi_i, \phi_j \rangle w^{m+s}(j)| + \hspace{-0.8cm}\sum_{\tiny j \in \{r+1, \ldots k\} \backslash I^{m+s}\cup \{i\}}\! \! \! \! \! \! \! \! \! \! \! \! |\langle \phi_i, \phi_j \rangle  w^{m+s}(j)|, \nonumber\\
&\overset{2}{\leq} (k-1) \mu (2k\mu |x_o(r+1)| + \ldots + (2k\mu)^{s}|x_o(r+1)| \nonumber \\
&+2(2k\mu)^{s+1}|x_o(r)|) + k\mu |x_o(r+1)| := \alpha_s.\nonumber
\end{align}
Equality $(1)$ is using the assumption that the first $r$ elements are in the active set at stage $m+s$. Inequality $(2)$ is also due to the assumptions of the induction and the fact that $w^{m+s}(j)= x_o(j)-x^{m+s}(j)$. \\
 At least one of the largest $k+1$ coefficients of $z$, corresponds to an element whose index is not in $\{1,2, \ldots k\}$, and the magnitude of this coefficient is less than $\alpha_s$. Therefore the threshold value is less than or equal to $\alpha_s$. Applying the soft thresholding to $z$ will at most add $\alpha_s$ to the distance of $z^{s+1}(i)$ and $x_o(i)$, and this completes the proof of the upper bound. The main thing that should be checked is whether the first $r$ elements will remain in the active set or not. For $i \in \{1,2 \ldots r\}$ we have,
 \begin{align}
	 |z^{m+s+1}(i)| & \geq |x_o(i)|- |z^{m+s+1}(i)-x_o(i)|, \nonumber \\
 	\geq  |x_o(i)| & -k\mu|x_o(r)|(1+2k\mu + \ldots +(2k\mu)^{s+1}) \nonumber \\
      -2  k\mu( & 2k\mu)^{s+1}|x_o(r)| \geq |x_o(i)|- \frac{|x_o(r)|}{2.05} \nonumber \\
     \geq  |x_o(r)|&- \frac{|x_o(r)|}{2.05}.
\end{align}
 If the sequence in the above expression is multiplied by $2$, the result will be a sequence in the form of the sequences mentioned in lemma \ref{seq_lem} for $\alpha=2k\mu$, $\beta=2$ and the last equality is based on that lemma. \\

If $i \notin \{1,2 \ldots k\}$,
 \begin{align}
	 |z^{m+s+1}(i)| &\leq  k\mu|x_o(r)|(1+2k\mu + \ldots +(2k\mu)^{s+1}) \nonumber \\
      &+2  k\mu( 2k\mu)^{s+1}|x_o(r)| \leq  \frac{|x_o(r)|}{2.05}. \nonumber
\end{align}
  Since $\min_{\{i: i \leq r\}}|z^{m+s+1}(i)| > \max_{\{i: i > k\}} |z^{m+s+1}(i)|$, the first $r$ elements remain in the active set. The base of the induction is also clear since it is the same as the assumptions of the lemma.
 \end{proof}

\begin{lem}\label{lem_soft2}
Suppose that $k \leq \frac{\mu^{-1}}{4.1}$, and $x_o(1), x_o(2), \ldots, x_o(r)$, $r \leq k$, are in the active set at the $m^{\rm th}$ step. Also, assume that $\frac{|x_o(r)|}{|x_o(r+1)|}\leq 2^{\ell_r-5}$. If
\begin{align}
|x^{m}(j)-x_o(j)| \leq 4k\mu |x_o(r)|,  \ \ \ \forall j \in I^{m}, \nonumber
\end{align}
then after $\ell_r$ steps $x_o(r+1)$ will get into the active set, and
\begin{align}
|x^{m+\ell_r+1}(j)-x_o(j)| \leq 4k\mu |x_o(r+1)|,  \ \ \ \forall j \in I^{m+ \ell_r+1}. \nonumber
\end{align}
\end{lem}
\begin{proof}
As before we try to find a bound for the error at time $m+\ell_r$. For $i \in \{1,2, \ldots, k\}$,

\begin{align}
|z^{m+\ell_r}(i)&-x_o(i)| \leq \frac{1}{2} |x_o(r+1)|( 2k\mu+ \ldots + (2k\mu)^{\ell_r}) \nonumber \\
                         & + (2k\mu)^{\ell_r+1}|x_o(r)| \leq  \frac{|x_o(r+1)|}{2.1} + \frac{|x_o(r+1)|}{64} \nonumber
\end{align}
and therefore for $i=r+1$,
\begin{align}
|z^{m+\ell_r}(r+1)| \geq &|x_o(r+1)|- |z^{m+\ell_r}(i)-x_o(i)| \geq \nonumber \\
& |x_o(r+1)| -\frac{|x_o(r+1)|}{2.1} - \frac{|x_o(r+1)|}{64}
\end{align}
Since $|z^{m+\ell_r}(r+1)| > \max_{\{i: k<i \}} |z^{m+\ell_r}(i)|$, the $r+1^{\rm th}$ element will get into the active set at this stage. On the other hand for any  $i \in I^{m+ \ell_r}$ we have $|x^{m+\ell_r}(i)-x_o(i)| \leq x_o(r+1)$. For the next stage of the algorithm we will have at most $2k$ non-zero $x^{m+ \ell_r}(i)- x_o(i)$ and absolute value of each of them is less than $|x_o(r+1)|$. Therefore $|z^{m+\ell_r+1}(i)- x_o(i)| \leq 2k\mu |x_o(r+1)|$ and after thresholding we have,  $|x^{m+\ell_r+1}(i)- x_o(i)| \leq 4k\mu |x_o(r+1)|$ for $i \in I^{m+\ell_r+1}$. \\
The base of the induction is also clear from the assumptions of this lemma and the proof is complete.
\end{proof}
For the IHT algorithm we proved that at the first step the first element will pass the threshold. Since the selection step of IST and IHT is exactly the same, we can claim that the same thing is true for IST, i.e. the largest magnitude coefficient will pass the threshold. Also, as we saw for IHT, the error was less than $k \mu |x_o(1)|$. Therefore, for the IST we have, $|x^1(j)-x_o(j)| < 2k\mu |x_o(1)|$. These bounds are even better than the bounds we need for \ref{lem_soft_1} and \ref{lem_soft2} and \ref{lem_soft3}. \\
The following lemma will explain what happens when the algorithm detects all the non-zero elements.

\begin{lem} \label{lem_soft3}
Suppose that $x_o(1), \ldots, x_o(k)$, are in the active set at the $m^{\rm th}$ step. Also assume that,
\begin{align}
|x^m(j)-x_o(j)| \leq 4k\mu|x_o(k)|. \nonumber
\end{align}
If $k\mu < \frac{1}{4.1}$, at stage $m+s$ all the elements remain in the active set and for every $j$ we will have,
\begin{align}
|z^{m+s}(j)-x_o(j)| \leq 2(2k\mu)^{s+1} |x_o(k)| \nonumber
\end{align}
\end{lem}
The proof of this lemma is very similar to the other lemmas and is omitted.

\begin{proof}[Outline of the proof of Theorem \ref{soft_thm}]
The proof is a simple induction by combining the above lemmas. Suppose that $x_o(1), x_o(2), \ldots, x_o(r)$ are already in the active set. According to Lemma \ref{lem_soft_1} all these terms will remain in the active set, and according to Lemma \ref{lem_soft2} after ${\ell_r} $ steps $x_o(r+1)$ will also get into the active set. In one more step, the error on each element gets smaller than $ 4k\mu |x_o(r+1)|$, and everything can be repeated. Although we have not mentioned the first step of the induction it is not difficult to see that step is also true and it is very similar to the first step of IHT. Finally when all the elements are in the active set lemma \ref{lem_soft3} tells us that the error goes to zero exponentially fast.
\end{proof}

\section{Discussion and Comparison With other Work}\label{discussion}

There is a huge amount of work on iterative thresholding algorithms, and we cannot mention all of them here; The interested reader is referred to \cite{MaDo09}. Most of these papers are dealing with a fixed threshold that does not depend on iteration. In that case, there are rigorous results that give sufficient conditions for the IST algorithm to converge to the solution of $\mathcal{P}_1$ \cite{DaDeDe04}, and for the IHT algorithm to a local minimum of $\mathcal{P}_0$ \cite{BlDa08}. The idea of choosing iteration dependent thresholds is also not new, and some simple variations were introduced in \cite{StElDo05}. The $k$ largest element thresholding policy was first introduced in \cite{NeTr08} and was first used for IHT in \cite{BlDa09}. It was also shown that if the $\Phi$ matrix satisfies restricted isometry property (RIP) of order $3k$, the IHT converges to the sparsest solution. There are some basic differences in our approach. First, we are dealing with deterministic settings, and in these settings RIP conditions they have provided are much weaker than ours ($k\mu < \frac{1}{3\sqrt{32}}$ compared to $k\mu <\frac{1}{3.1}$). Under these more general conditions, as we observed, the performance of IHT is not as simple as what is mentioned in \cite{BlDa09}, and it may not recover $x_o$ in just $k$ steps. But it will finally recover the sparsest signal and we give bounds on the number of iterations it needs to converge. Secondly, as discussed in the last section, our approach was easily adapted to IST, and can be adapted to the other types of thresholds. Moreover, our method gives us an ordering among  $\ell_1$, OMP, IHT and IST which may be useful for deciding on the choice of the algorithm. Finally there is another effort on analyzing the performance of IST by coherence that shows the possibility of success of such an algorithm at the first iteration \cite{HeGiTr06}. But this result does not have any conclusion about the next iterations of IST in case it does no recover all the non-zero elements at the first step.
\section{Conclusion}\label{conclusion}
In this paper, we analyzed iterative hard and soft thresholding, and proved that under certain conditions they work properly. These conditions are slightly weaker than their counterparts for $\ell_1$ and OMP. But these algorithms are very simple to implement and much faster than both convex relaxation and greedy methods, and they are much more desirable for large scale problems.
\section{Acknowledgement}
The author would like to thank David L. Donoho for helpful discussion and valuable suggestions on the early version of this manuscript. This work was partially supported by NSF DMS 05-05303.

\end{document}